\documentstyle[prl,aps,psfig,floats]{revtex}

\def\beq{\begin{equation}}
\def\eeq{\end{equation}}
\def\bea{\begin{eqnarray}}
\def\eea{\end{eqnarray}}

\def\ie{{i.e.}\ }
\def\eg{{e.g.}}
\def\eq{Eq.~}
\def\eqs{Eqs.~}
\def\fig{Fig.~}

\def\lam{\lambda}
\def\mpl{m_{\text{Pl}}}

\def\half{\frac{1}{2}}
\def\phidot{\dot{\phi}}

\def\flphi2{\langle\delta\phi^2\rangle}
\def\flchi2{\langle\delta\chi^2\rangle}
\def\Phid{\dot{\Phi}}
\def\delphi{\delta\phi}
\def\delchi{\delta\chi}
\def\delphid{\dot{\delta\phi}}

\def\sh{super-Hubble }
\def\ph{preheating }
\def\pr{parametric resonance }

\def\lex{Lyapunov exponent}

\begin{document}
\draft
\twocolumn[\hsize\textwidth\columnwidth\hsize\csname
@twocolumnfalse\endcsname
\title{Dynamical Chaos and the Growth of Cosmological Fluctuations}
\author{J. P. Zibin}
\address{Department of Physics and Astronomy, University of British Columbia, 
6224 Agricultural Road, Vancouver, BC V6T 1Z1 Canada}
\date{\today}
\maketitle

\begin{abstract}

   \ \ I demonstrate that instability in a system of homogeneous 
scalar fields leads to the growth of super-Hubble metric perturbations.  
This generalizes the result that parametric resonance can lead to 
the growth of cosmological perturbations.  Since dynamical chaos is 
common in multi-field quartically coupled systems, I argue that the 
evolution of the fields after inflation must be examined to determine 
whether the amplitude of cosmological metric perturbations 
is underestimated in the standard inflationary calculations.  
I illustrate this effect with a simple hybrid inflation model.

\end{abstract}
]

   {\em 1. Introduction.}--- A period of inflation has become the 
best candidate for the origin of 
structure in the universe.  Recent years have seen considerable 
interest in the process of reheating, which follows inflation and 
transforms the inflaton field's large potential energy into 
particle excitations of various fields.  A key development has 
been the understanding that \pr can lead to extremely rapid 
particle production in regions of $k$-space, a process known as 
preheating \cite{genph}.  If the \ph dynamics 
involves more than one scalar field, the resonance can extend to $k=0$, 
so that field perturbations which are \sh during preheating are amplified 
\cite{shph,zbs00}.  Large scale metric curvature 
perturbations can be amplified during \ph as well, which contradicts 
the usual assumption that they are constant on \sh scales.  
The growth of \sh metric perturbations is possible only when entropy 
modes are present --- single field perturbations are purely adiabatic 
and hence the usual conservation law applies \cite{wmll00,gwbm00}.  
These large scales are relevant to structure formation and the Cosmic 
Microwave Background (CMB), and hence it is crucial to understand 
this stage of the universe's evolution.

   In this Letter, I generalize previous studies of \pr in an attempt to 
understand under what conditions the growth of \sh metric perturbations 
is possible.  I point out a previously undiscussed general route 
for the amplification 
of \sh scalar field and metric curvature perturbations in multi-field 
models, namely through dynamical chaos in the background field evolution.  
(Note that ``dynamical chaos'' here refers to the {\em evolution} of 
a low-degree-of-freedom dynamical system of homogeneous fields, and 
should not be confused with ``chaotic inflation'', which 
refers to the insensitivity of certain inflationary models to the 
inflaton {\em initial conditions}.)  The idea is simple to state:  if 
the homogeneous background field dynamics is chaotic, then by definition 
small homogeneous perturbations of the backgrounds will grow 
exponentially on some characteristic time scale.  This implies the 
growth of large scale metric perturbations, which are the modes relevant 
to structure formation and the CMB.  Since dynamical chaos is common in 
nonlinear systems with two or more degrees of freedom (\eg\ in 
quartically coupled oscillators), I stress that the chaotic overproduction 
of \sh modes may rule out certain inflationary models.

   It is important to point out the distinction between \pr and dynamical 
chaos.  While both can result in exponential production of \sh 
modes, \pr can be described analytically while chaos cannot.  Parametric 
resonance techniques can be applied in the special case that the backgrounds 
(or conformally scaled backgrounds) undergo periodic motion.  Essentially, 
linear stability analysis is performed about the periodic orbit, 
and the presence of instability implies exponential growth of 
perturbations.  Dynamical chaos, on the other hand, involves 
instability to perturbations during extremely complex (and more general) 
evolution of the backgrounds.  The lack of periodic motion (and hence 
of parametric resonance) does {\em not} imply that perturbations 
cannot grow exponentially.

   There have been a number of studies of chaos in systems of homogeneous 
fields in cosmology, although apparently none have made the connection 
with the growth of \sh perturbations.  Easther and Maeda \cite{em99} 
studied the chaotic dynamics of a two-field hybrid inflation system 
during reheating, although they did not include metric perturbations.  
They found two effects:  the enhancement of defect formation, and a 
significant variation in the growth of the scale factor.  However, they 
claim that only scales $k\sim aH$ at the time of reheating 
will be affected.  Cornish and Levin \cite{cl96} studied a single 
field model, and followed the evolution for several ``cosmic cycles'' 
of bang and crunch.  Chaos is possible in such a simple system if 
gravity is important, as we will see below.  Latora and Bazeia 
\cite{lb00} studied a class of two-field quartically coupled systems 
which are chaotic in some regions of parameter space.

   {\em 2. Perturbation dynamics.}--- A general dynamical system consists 
of a set of $n$ (in general nonlinear) coupled first order ODEs
\beq
\dot{x_i} = F_i(x_j).
\label{eom}
\eeq
If we consider the state $x_i$ to be a function of the initial state 
$x_i^0$ (and time), we can write
\beq
\delta x_i = {\partial x_i(x_j^0) \over \partial x_j^0} \delta x_j^0.
\label{pert}
\eeq
In words, \eq(\ref{pert}) says that once we have obtained a ``background 
solution'' $x_i(x_j^0)$ to the equations of motion, we can obtain from 
it the evolution of a small perturbation simply by taking derivatives.  
(Of course in specific cases, it may be impossible to obtain the 
background solution in analytic form in order to implement this method.)

   The system (\ref{eom}) is said to exhibit dynamical chaos 
if the phase space is bounded and if a perturbation length 
$d(t) = (\delta x^i \delta x_i)^{1/2}$ grows exponentially with time, \ie
\beq
d(t) \sim d(t_0) e^{ht},
\label{lyap}
\eeq
for sufficiently small initial displacement $d(t_0)$ and sufficiently 
late $t$.  Here $h$ is known as the (largest) {\em Lyapunov 
exponent} \cite{o93}.  The requirement of a bounded phase space excludes 
trivially unstable systems, such as the inverted harmonic oscillator.  
Necessary conditions for dynamical chaos in the system (\ref{eom}) are that it 
contain nonlinear terms and that $n\geq3$ \cite{o93}, so that even very low 
dimensional systems can exhibit amplification of perturbations and 
the consequent extremely erratic background evolution.

   I now consider the case of a spatially flat Friedmann-Robertson-Walker 
background universe with $N$ minimally coupled real scalar fields $\phi_i$ 
described by the Lagrangian density
\beq
{\cal L} = \sqrt{-g}\left(\frac{1}{2}\sum_i\partial_\mu\phi_i\partial^\mu\phi_i
          - V(\phi_i)\right).
\eeq
The equations of motion for the homogeneous background scalar fields are
\beq
\ddot{\phi_i} + 3H\phidot_i + V_{,i} = 0,
\label{kg}
\eeq
where $V_{,i}=\partial V/\partial\phi_i$.  The background metric evolution 
is specified by the 0-0 Einstein equation,
\beq
H^2 = \frac{8\pi}{3\mpl^2}\left[\half\sum_i\phidot_i^2 + V(\phi_i)\right].
\label{fr}
\eeq
Note that on the assumption that $H$ does not change sign (which 
holds in realistic models where $H>0$), we can substitute the square 
root of \eq(\ref{fr}) into \eq(\ref{kg}).  The evolution equations 
for the set of functions $\{\phi_i,\phidot_i\}$ can then be written 
in the form of the dynamical system (\ref{eom}), with $n=2N$.  Thus 
with two or more nonlinearly coupled scalar fields, dynamical chaos 
is possible in the homogeneous background system.  (Indeed we can see 
that if we allow $H$ to change sign, as in bang-crunch scenarios, 
then we cannot eliminate $H$ from \eq(\ref{kg}).  In this case $n=2N+1$, 
and a single scalar field is sufficient for chaos, as was found 
in \cite{cl96}.)

   We can write the metric in the presence of scalar perturbations 
in the general form \cite{mfb92}
\bea
ds^2 &=& (1+2A)dt^2 - a^2(t)\{2B_{|i}dx^idt \nonumber\\
     & & - [(1-2\psi)\delta_{ij} + 2E_{|ij}]dx^idx^j\}.
\eea
Here $A$, $B$, $\psi$, and $E$ are scalar functions, and subscript $|i$ 
refers to the covariant derivative on the background constant time 
hypersurface.  Two of the four scalar functions may be determined 
by a choice of gauge, and the linear perturbation evolution equations 
will reflect that choice.  To obtain the long-wavelength limit of these 
equations, Sasaki and Tanaka \cite{st98} (see also \cite{u98}) noticed that 
it was sufficient to use \eq(\ref{pert}) (in a slightly different form).  
The issue of which gauge the perturbation equations are written in 
is tied to the choice of time variable used in the background equations 
\cite{st98}.  Explicitly, they found that to obtain the perturbed 
equations in a particular gauge, it is necessary to use a time 
variable which is {\em not perturbed} in that gauge.

   For example, in the synchronous gauge, $A=B=0$, the usual cosmological 
time $t$ itself remains unperturbed.  Thus 
\eq(\ref{pert}) applied to the background equations (\ref{kg}) gives
\beq
\ddot{\delphi_i} + 3\delta H \phidot_i + 3H\delphid + V_{,ij}\delphi_j = 0.
\eeq
With the identification
\beq
\delta H = \dot{\psi}
\eeq
we obtain the usual perturbation equation of motion in the synchronous 
gauge, in the limit $k\rightarrow0$.

   The importance of this result is that, if the homogeneous background 
dynamics is chaotic, the quantity $\delphi_i$ calculated above 
(which represents the homogeneous field perturbation in a particular 
gauge) {\em will} generically grow exponentially with time, as 
\eq(\ref{lyap}) indicates.  Since spatial derivative terms are 
negligible on \sh scales, this implies the exponential growth of \sh 
finite wavelength modes.  Finally, we can write the comoving curvature 
perturbation $\cal{R}$ by \cite{gwbm00}
\beq
{\cal R} = \psi + {H\over\dot{\sigma}}\delta\sigma,
\eeq
where $\sigma$ and $\delta\sigma$ are the field and perturbation 
components along the direction of the background trajectory.  Since the 
total length of the perturbation vector grows exponentially (unless 
the initial perturbation happens to be precisely adiabatic, corresponding 
to a perturbation in time), so must the component $\delta\sigma$, and 
hence so generically must $\cal{R}$.

   {\em 3. Hybrid inflation.}--- I can illustrate these 
ideas with a double scalar field model, which, as discussed above, 
is sufficiently complex for dynamical chaos to be possible.  A popular 
class of two-field inflationary models is {\em hybrid} inflation 
\cite{hybrid,l94}.  In these models inflation can be terminated 
by a symmetry breaking transition in one of the fields, and the 
subsequent oscillations can be chaotic \cite{em99}.  I will consider the 
potential
\beq
V(\phi,\chi) = {1\over4\lam}(M^2-\lam\chi^2)^2 + {1\over2}m^2\phi^2
             + {1\over2}g^2\phi^2\chi^2.
\eeq
Inflation occurs at large $\phi$, where the effective mass of the $\chi$ 
field, $m_\chi^2 = g^2\phi^2 +\lam\chi^2 - M^2$, is large and the $\chi$ 
field sits at the bottom of a potential valley at $\chi=0$.  Once 
the inflaton field drops below the critical value $\phi_c=M/g$, 
the mass-squared $m_\chi^2$ becomes 
negative (the potential valley becomes a ridge), and the 
fields undergo a symmetry breaking transition to one of the global 
minima at $\phi=0$, $\chi=\pm \chi_0$, where $\chi_0=M/\sqrt{\lam}$.  
I consider the ``vacuum-dominated'' regime, where the potential during 
the inflationary stage, $V(\phi) = M^4/(4\lam) + m^2\phi^2/2$, is 
dominated by the false vacuum energy term $M^4/(4\lam)$.  I also 
consider the case where the Hubble parameter at the critical point,
\beq
H_0^2 = {2\pi\over3\lam} {M^4\over\mpl^2},
\eeq
is much smaller than the oscillation frequencies about the global 
minima, which are $\overline{m}_\phi=gM/\sqrt{\lam}$ and 
$\overline{m}_\chi=\sqrt{2}M$ for small oscillations.  This ensures 
that the fields will oscillate very many times after the critical 
point is reached before Hubble damping is significant.

   Preheating has been studied in hybrid models for various parameter 
regimes in the absence of metric perturbations \cite{hybridph}.  
The behaviour of large-scale metric perturbations was studied in \cite{fb00}, 
where it was found that growth is possible on large scales.  Oscillations 
in hybrid inflation were found to be chaotic in \cite{em99}, although for 
a very different parameter regime than I examine here.  Also, the 
connection with the growth of large-scale perturbations was not made in 
\cite{em99}.

   Since my interest in this Letter is to establish a kinematical 
connection between dynamical chaos in the background fields and exponential 
growth of metric perturbations, I ignored the evolution of perturbations 
during the inflationary stage.  It is important to note that for the parameters 
I consider, the large $\chi$ mass during inflation implies damping of 
large scale perturbations during inflation.  Thus the question of 
whether the amplitude of metric perturbations produced in this model 
is consistent with the Cosmic Background Explorer normalization is 
not addressed here.  A careful analysis, following the evolution 
of all important scales during inflation and preheating, and including 
the effects of backreaction, is required \cite{zbs00}.

\begin{figure}
\centerline{\psfig{figure=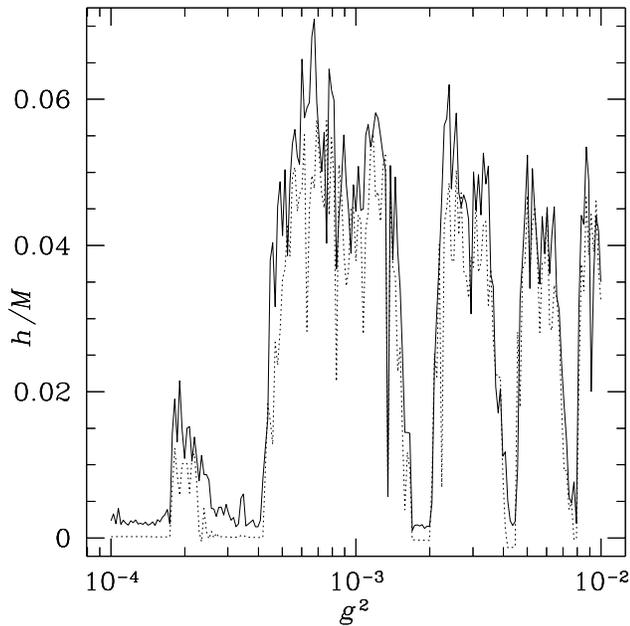,height=8.7cm}}
\caption{Largest \lex\ (solid line) for the homogeneous fields, and 
logarithmic growth rate of $\zeta_k$ (dotted line) for a scale 
$k/a=10^{-3}H_0$.  The homogeneous 
fields are oscillating about one of the hybrid model's global minima.  
The parameters are $M=10^{-8}\mpl$, $m=10^{-16}\mpl$, and $\lam=10^{-3}$.  
There is a clear correlation between the two curves.}
\label{lyapfig}
\end{figure}

   I considered the slice through parameter space specified by 
$M=10^{-8}\mpl$, $m=10^{-16}\mpl$, $\lam=10^{-3}$, and 
$g^2=10^{-2}$--$10^{-4}$.  These parameters give an amplitude of 
cosmological density perturbations of the order $10^{-5}$ according 
to the standard inflationary calculation \cite{l94}.  
I evolved the homogeneous background fields according to 
\eqs(\ref{kg}) with initial conditions $\phi(t_0)=0.999\phi_c$ 
and $\chi(t_0)=0.001\chi_0$.  I followed the evolution of 
the curvature perturbation on uniform-density hypersurfaces $\zeta_k$ 
(which coincides with $\cal{R}$ on large scales) for $k/a=10^{-3}H_0$ 
using the longitudinal gauge equations
\beq
\ddot{\delphi_i} + 3H\delphid_i + \frac{k^2}{a^2}\delphi_i
 + V_{,ij}\delphi_j = 4\Phid\phidot_i - 2V_{,i}\Phi,
\label{perkg}
\eeq
\beq
\Phid + H\Phi = \frac{4\pi}{\mpl^2}\phidot_i\delphi_i,
\label{perei2}
\eeq
\beq
\zeta_k = \Phi_k - \frac{H}{\dot{H}} \left( \Phid_k + H\Phi_k \right),
\label{zeta}
\eeq
where $\Phi_k$ is the longitudinal gauge metric perturbation.  
I calculated the largest \lex\ $h$ for the background evolution using 
\eq(\ref{lyap}).  The results, shown in \fig\ref{lyapfig}, indicate the 
presence of rich structure as $g^2$ is varied.  Regions of chaos with 
$h\simeq0.05M$ are interspersed with regular stability bands 
where $h\simeq0$.  The most prominent stability band is near the 
supersymmetric point $g^2/\lam=2$.  The correlation between 
the \lex\ and the growth rate of large scale metric perturbations 
is strong numerical evidence in support of my arguments.

\begin{figure}
\centerline{\psfig{figure=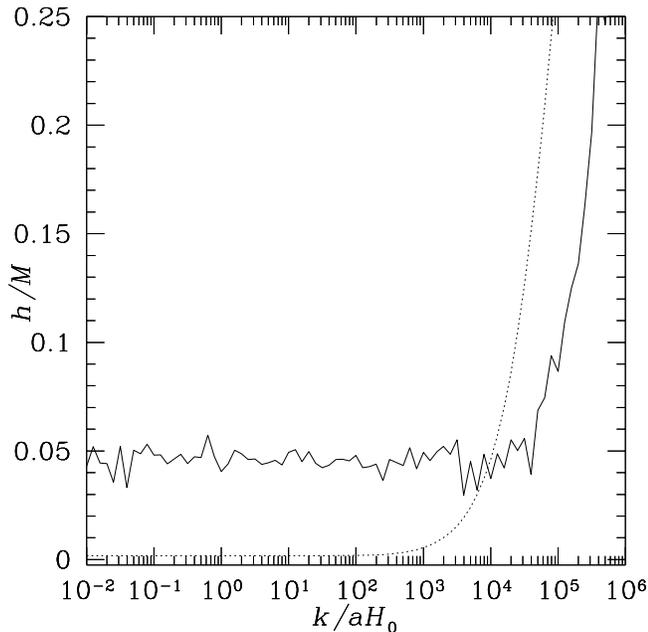,height=8.7cm}}
\caption{Logarithmic growth rate of $\delchi_k$ as a function of scale 
for the parameters of \fig\ref{lyapfig} and $g^2=10^{-3}$ (solid line) 
and for the trajectory constrained to $\phi=0$ (dotted line).}
\label{kdepfig}
\end{figure}

   The growth at large scales is not simply due to the negative 
mass-squared (or ``tachyonic'') instability near the potential ridge at 
$\chi=0$ \cite{fkl01}.  To demonstrate this, I plot in \fig\ref{kdepfig} 
the logarithmic growth rate of $\delchi_k$ perturbations as a function 
of scale $k$.  The solid line corresponds to the same parameters as 
in \fig\ref{lyapfig} (with $g^2=10^{-3}$), and shows a growth rate 
which is large at scales $k\simeq aM$ (due to the tachyonic instability), 
and approaches a constant (the chaotic \lex) as $k/(aH)\rightarrow0$.  
The dotted line shows the results for the same parameters except 
with the field trajectory constrained to $\phi=0$.  In this case, there 
is no large-scale growth (and no chaos) as expected for the effectively 
one-dimensional dynamics, while there 
is strong growth (due to the tachyonic instability) at small scales.  
Thus the two effects are distinct, since both trajectories 
pass through the tachyonic instability region, while only one exhibits 
growth on large scales.

   {\em 4. Summary.}--- I have demonstrated a general link between 
instability in scalar field background evolution and the growth of 
\sh metric perturbations.  In particular, dynamical chaos in the 
fields can drive the growth --- it is not necessary to have periodic 
motion and parametric resonance in the fields.  Since chaos is common 
in multi-field systems, it is important to examine the \sh evolution 
during \ph carefully, and also to follow the evolution during the 
inflationary period, in order to determine whether the model conflicts 
with CMB measurements.

   I wish to thank D. Scott, R. Brandenberger, W. Unruh, L. Kofman, and 
F. Finelli for helpful discussions.  This research was supported by the 
Natural Sciences and Engineering Research Council of Canada.

\end{document}